\documentclass[aps,prl,twocolumn,showpacs,preprintnumbers,superscriptaddress,amsmath]{revtex4}
\usepackage{txfonts}
\usepackage{amssymb}
\usepackage{graphicx}
\usepackage{dcolumn}
\usepackage{bm}

\begin{document}

\title{Coupling of higher-mode-light into a single sliver nanowire}
\author{Guo-Ping Guo}
\author{Rui Yang}
\author{Xi-Feng Ren\footnote{renxf@ustc.edu.cn}}
\author{Lu-Lu Wang}
\address{Key Laboratory of Quantum Information, University of Science and Technology of China, Hefei
230026, People's Republic of China}
\author{Hong-Yan Shi}
\author{Bo Hu}
\author{Shu-Hong Yu\footnote{shyu@ustc.edu.cn}}
\address{University of Science and Technology of China, Hefei
230026, People's Republic of China}
\author{Guang-Can Guo}
\address{Key Laboratory of Quantum Information, University of Science and Technology of China, Hefei
230026, People's Republic of China}
\begin{abstract}
Coupling of higher-order-mode light into a single sliver nanowire
and the degree of the coupling can be controlled by adjusting the
light polarization, showing that nanowire waveguide has no request
on the spatial mode of the input light. Photons with different
orbital angular momentums (OAM) are used to excite surface plasmons
of silver nanowires. The experiment indicates the propagating modes
of surface plasmons in nanowires were not the OAM eignenstates.

\end{abstract}
 \maketitle

Today, the major problem to increase the speed of microprocessors is
how to carry digital information from one end to the other. Optical
interconnectors can carry digital data much more than that of
electronic interconnectors, while fiber optical cables can not be
minimized to nanoscale due to the optical diffraction limit. To
solve this size-incompatibility problem, we may need to integrate
the optical elements on chip and fabricate them at the nanoscale.
One such proposal is surface plasmons, which are electromagnetic
waves that propagate along the surface of a conductor\cite{ozbay}.
Plasmonics, surface plasmon-based optics, has been demonstrated and
investigated intensively in nanoscale metallic hole
arrays\cite{Ebbesen98,Moreno,Alt}, metallic
waveguides\cite{Pile,Bozhevo,Lamp}, and metallic
nanowires\cite{Dickson,Graff,Ditlbacher,Sanders,Knight,Pyayt} in
recent years.

Among the various kinds of plasmonics waveguides, sliver nanowires
have some unique properties that make them particularly attractive,
such as low propagation loss due to their smooth surface and
scattering of plasmons to photons only at their sharp ends. Since
the momentums of the photons and plasmons are different, it is a
challenge to couple free-space light into plasmon waveguides
efficiently. The typical methods for plasmon excitation include
grating coupling, prism coupling and focusing of light onto one end
of the nanowire with a microscope objective. Nanoparticle
antenna-based approach is also proved as an effective way for
optimizing plasmon coupling into nanowires\cite{Knight}, which
realizes direct coupling into straight, continuous nanowires by
using a nanoparticle as an antenna. Recently, polymer waveguides are
used to couple light into several nanowires
simultaneously\cite{Pyayt} as well, aiming at providing light to a
number of nanoscale devices in the future integrated photonic
circuits.

\begin{figure}
\includegraphics[width=7.0cm]{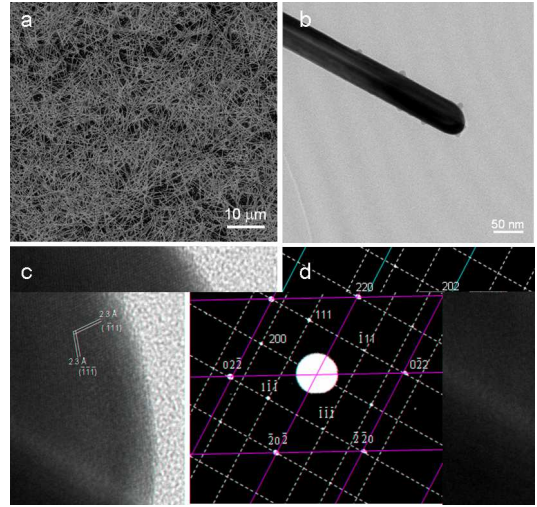}
\caption{(color online)(a) SEM image of silver nanowires. (b) TEM
image taken on the end of an individual silver nanowire. (c) HRTEM
image of the singe nanowire shown in (b). (d) SAED pattern taken
from an individual nanowire.}
\end{figure}
Because the former researches about the nanowires always concentrate
on using Gaussian mode light to excite surface plasmons, here we
discuss whether surface plamons can be launched by other
higher-order-mode light. We focus laser beam with different orbital
angular momentums (OAM) on one end of a nanowire and observe
scattering light from the other end. Surface plasmons are launched
not only by Gaussian mode light but also by higher-order-mode light.
The coupling strength over light polarization is also studied for
higher-order-mode light and gives the similar results with the case
of Gaussian mode. The output intensity increases linearly with the
input intensity rising, and is independent of the spatial mode of
the input light.

Ag nanowires were synthesized through a polyol process in a mixture
of ethylene glycol (EG) and poly (vinyl pyrrolidone) (PVP) at a
certain temperature, which was very similar as the previous
report\cite{tao,jiang,korte}. Scanning electron micrograph (SEM)
image in Fig.1a shows that all the nanowires are straight and have
uniform diameters that vary from 60 to 100nm and lengths from 10 to
40$\mu m$. A typical nanowire with diameter of 60 nm is shown in
Fig. 1b. High resolution TEM image in Fig. 1c shows a lattice
spacing of 0.23 nm, corresponding to those of
$(\bar{1}\bar{1}\bar{1})$ and $(\bar{1}11)$respectively. Electron
diffraction pattern taken the individual nanowire can be indexed as
two parallel zone axes, i.e. $[01\bar{1}]$ and
$[1\bar{1}\bar{1}]$(Fig. 1d). Based on the analysis, the nanowire
axis is along $[100]$.

The mode of the input light is determined by its OAM. It is known
that photons have both spin angular momentum and OAM. The light
fields of photons with OAM can be described by means of
Laguerre-Gaussian ($LG_p^l$) modes with two indices $p$ and
$l$\cite{Allen92}. The $p$ index identifies the number of radial
nodes observed in the transversal plane and the $l$ index describes
the number of the $2\pi$-phase shifts along a closed path around the
beam center. If the mode function is a pure LG mode with winding
number $l$ , then every photon of this beam carries an OAM of
$l\hbar $. This corresponds to an eigenstate of the OAM operator
with eigenvalue $l\hbar $\cite{Allen92}. For the sake of
simplification, here we just consider the cases for $p=0$. When
$l=0$, the light is in the general Gaussian mode, while when $l\neq
0$, the energy distribution of light likes a doughnut due to their
helical wavefronts (see inset of Fig. 2). We usually use computer
generated holograms (CGHs)\cite{ArltJMO,VaziriJOB} to change the
winding number of LG mode light. It is a kind of transmission
holograms. Inset of Fig. 1. shows part of a typical CGH($n=+1$) with
a fork in the center. Corresponding to the diffraction order $m$,
the $l$ fork hologram can change the winding number of the input
beam by $\Delta l_m=m*n$. In our experiment, we use the first order
diffraction light ($m=+1$) and the efficiencies of  our CGHs are all
about $40\%$. The Gaussian mode light can be identified using
mono-mode fibers in connection with avalanche detectors. All other
modes light have a larger spatial extension, and therefore cannot be
coupled into the single-mode fiber efficiently.

\begin{figure}
\includegraphics[width=8.0cm]{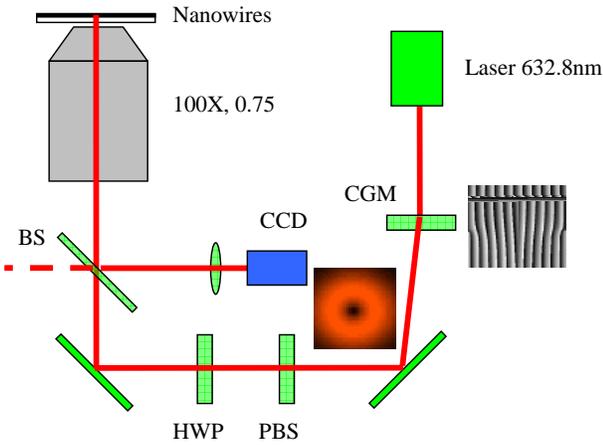}
\caption{(color online)Sketch illustration of the experimental
setup. The OAM of the laser (wavelength 632.8nm) was controlled by a
CGH, while the polarization was controlled by a PBS followed by a
HWP. The polarized laser beam was focused on one end of a nanowire
using a 100X objective lens (Zeiss, NA=0.75). The sample was moved
by a three dimensional piezo-electric stage. Scattering light was
recorded by a CCD camera after a microscope objective. Inset are
pictures of a typical CGH ($n=1$) and the energy distribution of the
produced light.}
\end{figure}

The experimental setup was shown in Fig. 2. The wavelength of the
laser beam was 632.8 nm, which was much bigger than the diameter of
the nanowires (about 100 nm). The OAM of the laser was controlled by
a CGH, while the polarization was controlled by a polarization beam
splitter (PBS, working wavelength 632.8 nm) followed by a half wave
plate (HWP, working wavelength 632.8 nm). Rotating the HWP allowed
us to investigate the relation between the coupling efficiency and
the polarization of light. The polarized laser beam was directed
into the microscope and focused on one end of a nanowire with the
light diameter about $5.5\mu m$ using a 100X objective lens (Zeiss,
NA=0.75). The sample was moved by a three dimensional piezo-electric
stage (Physik Instrumente Co., Ltd. NanoCube XYZ Piezo Stage).
Scattering light from the nanowire was reflected by a beam splitter
(BS, 50/50) and recorded by a CCD camera.

\begin{figure}
\includegraphics[width=8.0cm]{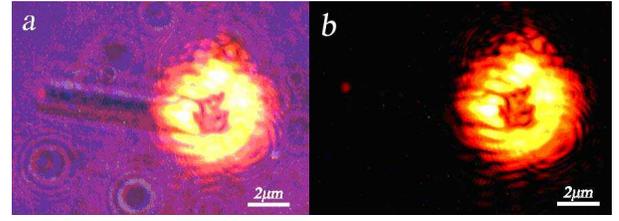}
\caption{(color online)Higher-order-mode light ($l=2$) was focused
on one end of a nanowire (length $9.3\mu m$) and the emission was
observed from the other end clearly, which verified that the
higher-mode-light can also be transmitted by the sliver nanowire.}
\end{figure}

The momentum of the propagating plasmon($k_{sp}$) is larger than
that of the incoming photon($k_{ph}$), there needs an additional
wavevector($\Delta k$) to sustain the momentum conservation
condition. Surface plasmons in nanowires can be excited when the
symmetry were broken, for example, at the ends and sharp
bends\cite{Dickson,Graff,Ditlbacher,Sanders}, because an extra
wavevector ($\Delta k_{scatter}$) is provided according to the
scattering mechanism at this situation. It has been proved that
surface plasmons can propagated along the length of nanowires when
they were excited by Gaussian mode light, even the diameter of
nanowires were much smaller than the wavelength of light. In our
experiment, higher mode lights ($l=1$ and $2$) were focused on one
end of a nanowire (length $9.3\mu m$ ) and the emission was observed
from the other end clearly, which verified that the
higher-mode-light can also be transmitted by the sliver nanowire
(Fig. 3). It is noted that the energy distribution of the output
light was not same with the input light which has a null hole in
center. This phenomenon is different from the cases of extraordinary
optical transmission through nano-hole structures, where the OAM
eigenstates can be preserved\cite{ren06,wang}. A potential
explanation is that the propagation modes of surface plasmons in
nanowires are not the eigenmodes of OAM states, like the case of
multi-mode optical fiber.
\begin{figure}
\includegraphics[width=8.0cm]{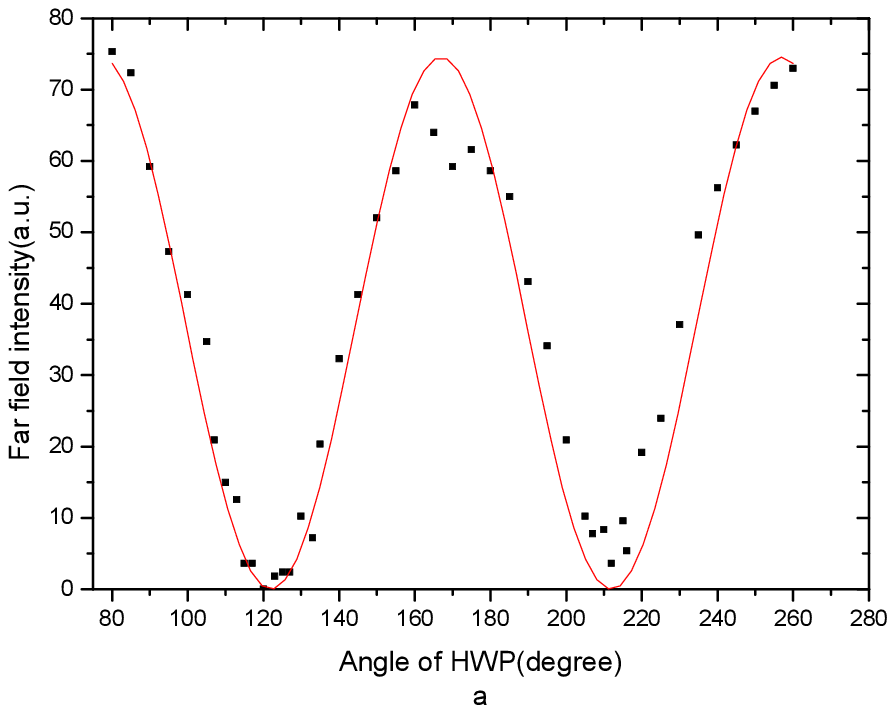}
\includegraphics[width=8.0cm]{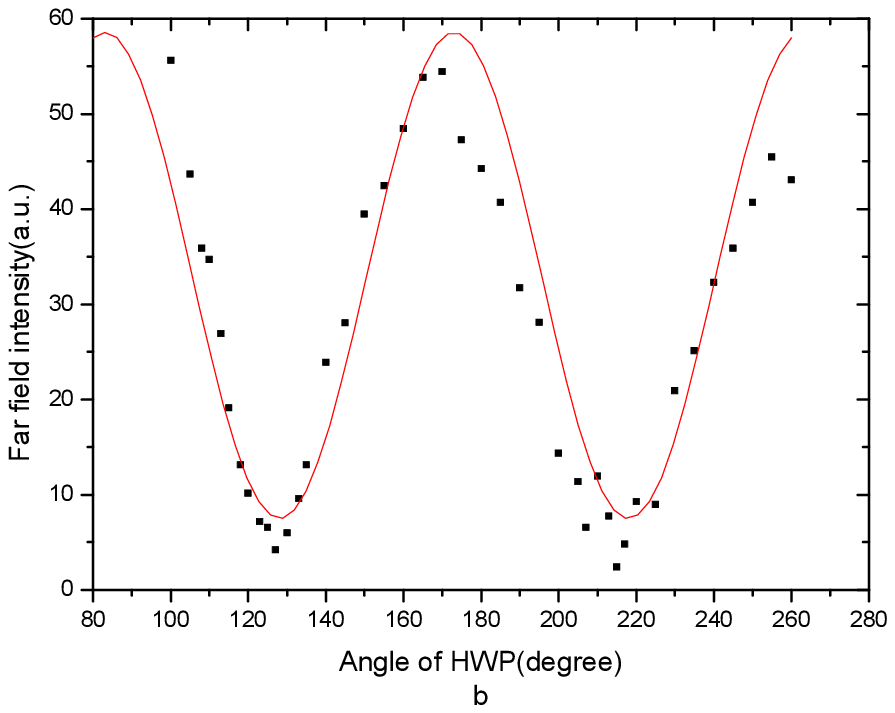}
\caption{(color online)Polarization dependence of coupling
efficiency at nanowire end. (a) The Gaussian mode light was focused
on the end of nanowire. (b) The higher-order-mode light ($l=2$) was
focused on the end of nanowire. The two cases give the similar
curve.}
\end{figure}

The coupling strength was measured by changing the polarization of
the input light. For each polarization, the emission intensity was
determined by averaging the four brightest pixels at the other end
of the nanowire in the CCD images. It changed with the polarization
of input light for the different coupling efficiencies. Fig. 4 shows
the relationship between the coupling strength and the polarization
of the input light. The far-field emission curves as a function of
polarization angle was approximately in accord with the theoretical
prediction (cosine or sine function)\cite{Sanders,Knight}. As a
comparison, the case of Gaussian mode light was observed and gave
the similar curve.

The end of the nanowire was also moved from one edge to the other of
the laser spot (which has a diameter about $5.5\mu m$) to give the
relationship between the input intensity of laser beam and the
emission intensity from the end of the nanowire. The results were
measured for the cases of Gaussian mode light and higher-order-mode
light ($l=2$), as shown in Fig. 5, which showed that the emission
intensity increased linearly with the pump intensity and was almost
independent of the spatial mode of the input light.

\begin{figure}
\includegraphics[width=8.0cm]{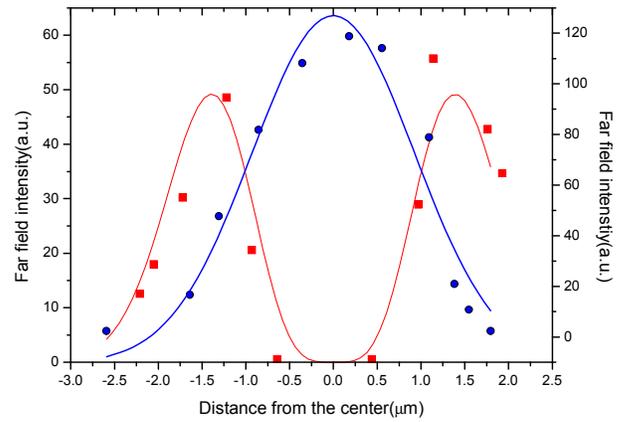}
\caption{(color online)Relationship between the input intensity of
laser beam and the emission intensity from the nanowire. The dots
are experimental results and the lines come from theoretical
calculations. The Gaussian mode light (blue round dots) and the
higher-order-mode light (red square dots) were focused on the end of
a nanowire. In both cases, the emission intensity changed linearly
with the input intensity.}
\end{figure}

In conclusion, we experimentally demonstrate that higher-order-mode
light can also excite surface plasmons in sliver nanowires. The
surface plasmons can propagate along the nanowire and scatter back
to photons at the other end. The coupling strength is correlated
with the polarization of input light, as the same as the case of
Gaussian mode light. The OAM eigenstates are not the propagating
modes of surface plasmons in nanowires. These results may give us
more hints to the understanding of the waveguide properties of
sliver nanowires.

This work was supported by National Fundamental Research Program
(Nos. 2006CB921900, 2005CB623601), National Natural Science
Foundation of China (Nos. 10604052, 50732006, 20621061, 20671085),
Chinese Academy of Sciences International Partnership Project, the
Partner Group of the CAS-MPG and Natural Science Foundation of Anhui
Province(Grants No. 090412053)..

\end{document}